\documentclass[doublecol,reprint]{epl2} 
\usepackage{amsmath,amssymb}
\usepackage{verbatim}
\usepackage{graphicx}
\usepackage{color}
\usepackage[colorlinks,bookmarks=false,citecolor=blue,
linkcolor=red,urlcolor=blue]{hyperref}
\usepackage{times}
\usepackage{soul}

\def \beq{\begin{equation}}
\def \eeq{\end{equation}}
\def \barray{\begin{eqnarray}}
\def \earray{\end{eqnarray}}

\begin{document}

\author{Vincenzo Alba\inst{1,2} \and Pasquale Calabrese\inst{2}}
\institute{
\inst{1}Delta Institute for Theoretical Physics, University of Amsterdam,
Science Park 904, 1098 XH Amsterdam, the Netherlands. \\
\inst{2} International School for Advanced Studies (SISSA) and INFN,
Via Bonomea 265, 34136, Trieste, Italy, 
International Centre for Theoretical Physics (ICTP), I-34151, Trieste, Italy. 
}

\date{\today}

\title{Quantum information dynamics in multipartite integrable systems}  

\abstract{
In a non-equilibrium many-body system, the quantum information dynamics between non-complementary regions is a
crucial feature to understand the local relaxation towards statistical ensembles. 
Unfortunately, its characterization is a formidable task, as non-complementary parts are generally in a mixed state. 
We show that for integrable systems, this quantum information dynamics can be quantitatively 
understood within the quasiparticle picture for the entanglement spreading. 
Precisely, we provide an exact prediction for the time evolution  of the logarithmic 
negativity after a quench. 
In the space-time scaling limit of long times 
and large subsystems, the negativity becomes proportional to 
the R\'enyi mutual information with R\'enyi index $\alpha=1/2$. 
We provide robust numerical evidence for the validity of our results 
for free-fermion and free-boson models, but our 
framework applies to any interacting integrable system.}

\maketitle

An isolated homogeneous quantum system evolving from a non-equilibrium pure state {\it locally} 
relaxes, for large time,  to a statistical ensemble \cite{ge-15,cem-16,ef-16}.
This relaxation is elucidated by the quantum information dynamics of a bipartition.
Indeed, focusing for simplicity on a one-dimensional system,  the entanglement entropy of an arbitrary compact subsystem 
of length $\ell$ initially grows  linearly in time and later saturates to a value that is extensive in $\ell$ \cite{cc-05,CC:review}. 
This saturation value is  the thermodynamic entropy of the ensemble describing the 
local behavior of the system  \cite{dls-13,collura-2014,bam-15,kauf,alba-2016,nahum-17,nahum-18,zn-18,alba-2018,p-18,nwfs-18} (which is either 
a thermal or a generalized Gibbs ensemble (GGE) depending on the conserved local charges, see, 
e.g., \cite{D91,S94,R08,dalessio-2015,rigol-14,rigol-2007,barthel-2008,cramer-2008,cef-12,fe-13b,sotiriadis-2014,ilievski-2015a,langen-15,vidmar-2016}). 
The linear growth of the entanglement entropy is  the main limitation 
to simulate the time evolution of isolated quantum systems with tensor network techniques \cite{swvc-08,pv-08,rev0,d-17,lpb-18}.  
However, this state-of-the-art may appear odd, as for very large time, the system is locally in a statistical ensemble 
which has very little entanglement and most of its entropy is just of statistical nature (implying, for instance, that the 
finite temperature properties are easily accessed by tensor networks \cite{uli-2011,hlbazp-18}). 
Thus, the large global entanglement, that is the main obstacle for tensor networks, locally arises mainly from classical fluctuations. 
It would be then highly desirable to devise new techniques that can effectively capture the local 
dynamics of a subsystem without worrying about the never ending growth of the entropy of the entire system (see e.g. \cite{mcpc-10,hlbazp-18,peat-14}). 
The entanglement between  finite non-complementary regions is expected to provide a figure of merit for 
 the effectiveness of an algorithm of this kind.   
We argued that this entanglement should be small for short and large times, but what about intermediate times?
There is little evidence \cite{lpb-18,d-17,hlbazp-18,jhn-18,ctc-14}  showing the presence of an ``entanglement barrier'' that might grow with subsystem size, but its
quantification  is a daunting task, because the two parts are in a mixed state. 
Therefore, in the following we will focus on the entanglement negativity which is a good entanglement measure for mixed states.
The calculation  of  the negativity is highly non-trivial  and results for global quantum quenches 
to the date are limited to some simple quenches in conformal field theories \cite{ctc-14} (see anyhow Refs. \cite{hoogeveen-2015,eisler-2014,
wen-2015,gh-18} for other non-equilibrium protocols). 
Here we will show that, very surprisingly, the quasiparticle picture of Refs. \cite{cc-05,cc-07} can be adapted to exactly quantify the negativity in integrable systems
after a global quench from low-entangled initial states, in the space-time scaling limit of long times and large subsystems with their ratio fixed.
We show that in this limit the negativity becomes proportional to the R\'enyi mutual information (with $\alpha=1/2$) which generically is only an upper bound for the entanglement.
We are able to test this result only for free bosonic and fermionic theories, but we expect to be valid for all integrable systems.

\begin{figure}
\centerline{\includegraphics[width=1\linewidth]{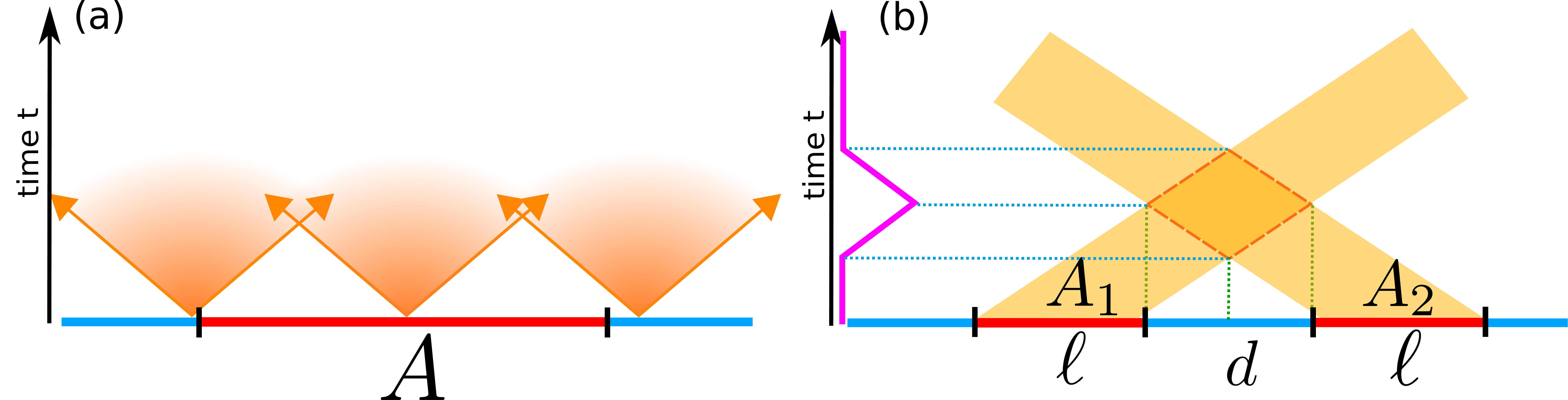}}
\caption{ Quasiparticle picture for the evolution of the negativity 
 after a quench in integrable systems. 
 (a) In the bipartite setting of $A$ and its complement $\bar A$, entangled quasiparticle pairs are produced 
 uniformly  and move ballistically (shaded cones). 
 (b) Tripartite pure state:  $A=A_1\cup A_2$, with $A_{1(2)}$ 
 two intervals of length $\ell$ at distance $d$. The negativity 
 between $A_1$ and $A_2$ is proportional to the horizontal width of the 
 intersection between the two shaded areas at a given time. 
 The resulting negativity  dynamics for the case of quasiparticles with perfect linear 
 dispersion is shown on the left side (purple lines).
}
\label{fig0}
\end{figure}

{\it The  logarithmic negativity} ${\cal E}_{A_1:A_2}$ ~\cite{peres-1996,zycz-1998,zycz-1999,lee-2000,vidal-2002,plenio-2005}  is defined as 
\begin{equation}
\label{neg}
{\cal E}_{A_1:A_2}\equiv \ln||\rho_A^{T_2}||_1=\ln\textrm{Tr}|\rho^{T_2}_{A_1\cup A_2}|. 
\end{equation}
Here $\rho_A^{T_2}$ is the partial transpose of the reduced density matrix  $\rho_{A}$
defined as $\langle\varphi_1\varphi_2 |\rho^{T_2}_A|\varphi'_1\varphi'_2\rangle\equiv\langle
\varphi_1\varphi_2'|\rho_A|\varphi'_1\varphi_2\rangle$, with 
$\{\varphi_1\}$ and $\{\varphi_2\}$ two bases for $A_1$ and $A_2$, respectively. 
The negativity is a good measure of entanglement between $A_1$ and $A_2$ 
(i.e. it is an entanglement monotone~\cite{vidal-2002,plenio-2005})
also when $A_1\cup A_2$ is in a mixed state. 
Another quantity we will consider is the R\'enyi mutual information 
\begin{equation}
\label{mi}
I^{\scriptscriptstyle(\alpha)}_{A_1:A_2}
\equiv S^{\scriptscriptstyle(\alpha)}_{A_1}+S^{\scriptscriptstyle(\alpha)}_{A_2}-
S^{\scriptscriptstyle{(\alpha)}}_{A_1\cup A_2},
\end{equation}
where $S^{\scriptscriptstyle(\alpha)}_{A_i}\equiv 1/(1-\alpha)\textrm{Tr}
\rho_{A_i}^\alpha$ are the R\'enyi entropies and the index $\alpha$ is a real number. 
The mutual information is an upper bound for the entanglement 
between $A_1$ and $A_2$ since it is sensitive to the total correlations, both quantum and classical.
The condition ${\cal E}_{A_1:A_2}=0$ is only necessary (not sufficient) for the absence of mutual entanglement.

Recently, the negativity became a useful tool to characterize universal aspects of 
quantum many-body systems~\cite{hannu-2008, mrpr-09,hannu-2010,calabrese-2012,cct-neg-long,calabrese-2013,calabrese-2015,kpp-14,ruggiero-2016,rr-15,fournier-2015,bc-16,
lee-2013,castelnovo-2013,hart-2018,java-2018,bayat-2012,bayat-2014,abab-16,wen-2016,wen-2016a,ruggiero-2016a,glen,alba-e,gs-18,dct-15,gbbs-17,csg-18,sherman-2016,lu-2018}
also because it can be computed in tensor network simulations~\cite{hannu-2008,calabrese-2013,ruggiero-2016}. 
The negativity~\eqref{neg} can be obtained for free-bosonic models in arbitrary dimension by correlation matrix methods~\cite{audenaert-2002,eisler-2016,dct-16}, 
but not for free fermions~\cite{eisler-2014a,coser-2015,ctc-16,chang-2016,hw-16,shapourian-2016,ssr-16,eez-16}.
Hence,  an alternative entanglement monotone which is effectively calculable using standard free-fermion 
techniques has been introduced ~\cite{shapourian-2016,ssr-16,shiozaki-2017,shiozaki-2018,shapourian-2018,shapourian-2018a} and 
it is also an upper bound for the original negativity~\eqref{neg}~\cite{eez-16}.
In the following, we denote as ${\cal E}^{\scriptscriptstyle(f)}_{A_1:A_2}$ this fermionic negativity (see Ref.~\cite{shapourian-2016} and below 
for its definition), and ${\cal E}^{\scriptscriptstyle(b)}_{A_1:A_2}$ the one in~\eqref{neg}.

{\it The quasiparticle picture} for the entanglement spreading \cite{cc-05} is illustrated in Fig.~\ref{fig0}. 
Entangled quasiparticle pairs are generated uniformly at $t=0$ by  the quench (see Fig.~\ref{fig0} (a)). 
Quasiparticles produced at the same point in space are 
entangled, whereas quasiparticles created far apart are 
incoherent. At a generic time $t$, the entanglement 
between a subsystem $A$ (see Fig.~\ref{fig0} (a)) and 
the rest is proportional to the total number of 
quasiparticles created at the same point at $t=0$ 
and shared between $A$ and its complement at time $t$. 
For integrable models, it has been shown~\cite{alba-2016,alba-2018} 
that, by combining the quasiparticle picture with integrability-based knowledge of the stationary state \cite{ce-13}, 
it is possible to accurately describe  the 
post-quench evolution of the von Neumann entropy and 
the mutual information, in the space-time scaling limit.
Two crucial features emerge. 
First, the stationary density of entanglement entropy is the same as that of the 
{\it thermodynamic} entropy, i.e. the GGE one.
Second, the entangling quasiparticles are the low-lying excitations around the stationary state. 
This picture has been extended and applied to describe the steady-state R\'enyi entropies~\cite{AlCa17,AlCa17b,mestyan-2018},  quenches from 
piecewise homogeneous initial states~\cite{transport,transport2}, and  few other non-equilibrium situations \cite{fnr-17,mkz-17,eh-cft,btc-18,bc-18}. 

It has been already argued in Ref.~\cite{ctc-14}--in the context of conformal field theories, i.e., theories with a unique velocity $v$--that 
the quasiparticle description can be applied to the entanglement negativity.
As the negativity measures the {\it mutual} entanglement between the intervals, the natural assumption is that 
at a generic time $t$, it must be proportional to the total number of entangled pairs shared between $A_1$ and $A_2$.
Hence, the picture for two intervals $A_1$ and $A_2$ of equal length $\ell$ at distance $d$ is illustrated in Fig.~\ref{fig0} (b). 
For  a single type of quasiparticles with fixed velocity $v$,  the number of shared particles
is proportional to the width at the time $t$ of the intersection between the two shaded areas starting from $A_1\cup A_2$ ~\cite{ctc-14} (see Fig.~\ref{fig0} (b)). 
The emission region of the pairs is given by the projection of this width onto the $t=0$ axis (vertical dotted lines in the Figure). 
The resulting negativity dynamics has the triangular form depicted as a function of time on the left side:
at short times, i.e., for $t<d/(2v)$, 
the negativity vanishes;
for $d/(2v)\le t\le(d+\ell)/(2v)$ there is a linear increase, followed by a linear decrease with the 
same (in absolute value) slope until $t\le(d+2\ell)/(2v)$, when the negativity vanishes and stays zero for all larger times. 
In formulas one has 
${\cal E}_{A_1:A_2}^{(f/b)}= \varepsilon^{(f/b)}[\textrm{max}(d/2,2|v|t)+\textrm{max}((d+4\ell)/2,2|v|t)
-2\textrm{max}((d+2\ell)/2,2|v|t)]$, where $\varepsilon^{\scriptscriptstyle(f/b)}$, 
is related to the rate of production of quasiparticles and to their contribution to the negativity. 

%
\begin{figure*}[t]
\includegraphics*[width=0.98\linewidth]{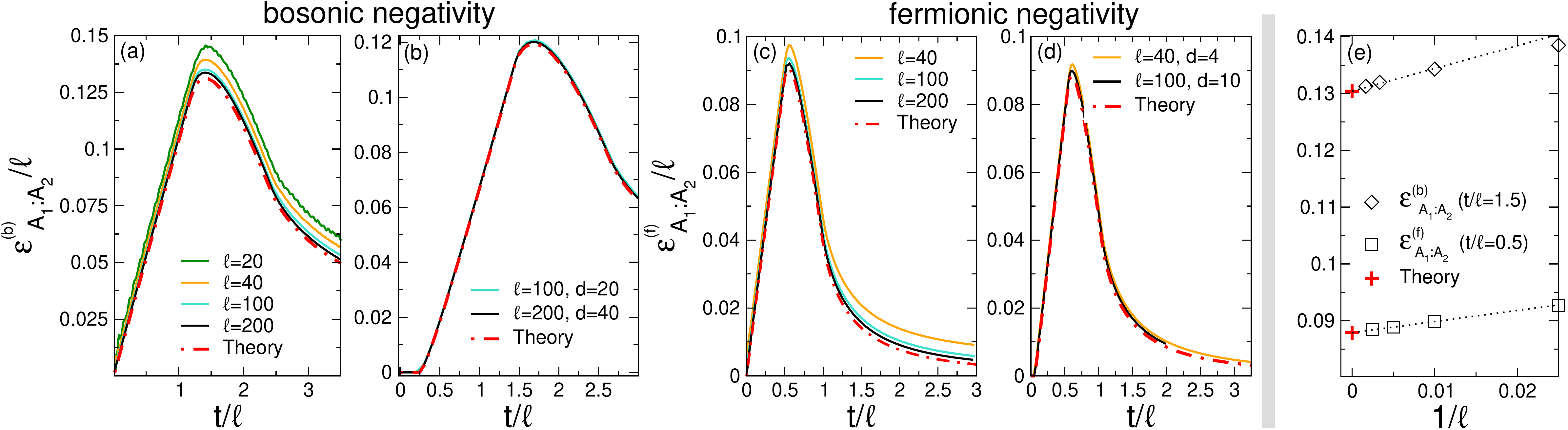}
\caption{ Out-of-equilibrium dynamics of the negativity after 
 a quench: Quasiparticle picture versus numerical results. 
 (a,b). Bosonic negativity ${\cal E}_{A_1:A_2}^{\scriptscriptstyle(b)}$ 
 after the mass quench from $m_0=1$ to $m=2$ in the harmonic chain:  
 ${\cal E}^{\scriptscriptstyle(b)}_{A_1:A_2}/\ell$, plotted versus 
 rescaled time $t/\ell$, with $\ell$ the intervals length, 
 for adjacent (in (a)) and disjoint intervals at distance $d$ 
 (in (b)). In (b) we choose configurations with 
 fixed ratio $\ell/d=5$. Continuous lines are results for 
 different $\ell$. The dashed-dotted line is the analytic 
 result in the scaling limit. (c,d). Same as in (a,b) for 
 the fermionic negativity 
 ${\cal E}^{\scriptscriptstyle(f)}_{A_1:A_2}$ after a quench in 
 the Ising chain (with $h_0=10$ and $h=2$). In (d) we fix $\ell/d=10$. 
 (e) Finite-size scaling corrections for adjacent intervals. Square 
 and diamond symbols are for the fermionic and bosonic negativity 
 at fixed $t/\ell=0.5$ and $t/\ell=1.5$, respectively. 
 Quenches are the same as in (a) and (c). 
 The crosses are the theoretical results in the thermodynamic limit. 
 Dotted lines are linear fits. 
}
\label{fig1}
\end{figure*}
%
For generic integrable models 
the quasiparticles of momentum $k$ have a nontrivial dispersion $e(k)$. 
Generically both the velocities $v(k)\equiv de(k)/dk$ and $\varepsilon^{\scriptscriptstyle(f/b)}(k)$ depend on $k$. 
The negativity is obtained by integrating over the independent quasiparticle contributions, to obtain 
\begin{multline}
\label{quasi-1}
{\cal E}_{A_1:A_2}^{(f/b)}=
\int
\frac{dk}{2\pi}\varepsilon^{(f/b)}(k)[\textrm{max}(d/2,2|v|t)\\+\textrm{max}((d+4\ell)/2,2|v|t)
-2\textrm{max}((d+2\ell)/2,2|v|t)], 
\end{multline}
Here we assumed the existence of a single species of quasiparticles, but in the most general case it is sufficient to sum over the independent species as 
for the entanglement entropy \cite{alba-2016,alba-2018}.
Notice that Eq.~\eqref{quasi-1} is the same as for the R\'enyi mutual information (of any index)
provided that one replaces $\varepsilon^{\scriptscriptstyle(f/b)}$ with the density of R\'enyi entropy. 

As it stands, Eq.~\eqref{quasi-1} describes the qualitative evolution of the negativity after a quench. 
To make it quantitative, we should fix both the velocities $v(k)$ and $\varepsilon^{\scriptscriptstyle(f/b)}(k)$.
For the former there is no difference compared to the entanglement entropy being $v(k)$ the velocities of the excitations around the stationary GGE.
For free models $v(k)$ does not depend on the quench, while for interacting models, $v(k)$ depends on the GGE and 
can be calculated using well-known Bethe ansatz methods~\cite{bonnes-2014}.
Fixing $\varepsilon^{\scriptscriptstyle(f/b)}(k)$ is non trivial and to the date is the main obstacle to make Eq. \eqref{quasi-1} predictive. 

Our crucial observation is that the single particle contribution  $\varepsilon^{\scriptscriptstyle(f/b)}(k)$ may be fixed by considering the negativity of a single interval, 
i.e. a bipartite pure state with $A_2=\bar A_1$.
In this case, the negativity equals the R\'enyi entropy with $\alpha=1/2$ both for bosons \cite{vidal-2002} and fermions \cite{eez-16}:
\begin{equation}
\label{bip}
{\cal E}^{(b)}_{A:\bar A}={\cal E}^{(f)}_{A:\bar A}
=S^{(1/2)}_A=S^{(1/2)}_{\bar A}. 
\end{equation}
Eq.~\eqref{bip} implies that $\varepsilon^{\scriptscriptstyle(b/f)}(k)$ coincides with the analogous quantity for the  R\'enyi entropy which can be extracted from the GGE
thermodynamic entropy   
$S^{\scriptscriptstyle(1/2)}_{\textrm{GGE}}$.
In terms of the mutual information, one has 
\begin{equation}
\label{n-mi}
{\cal E}_{A_1:A_2}^{(b/f)}=\frac{I_{A_1:A_2}^{(1/2)}}{2},
\end{equation}
as it  is clear from~\eqref{mi} by using that $A_1\cup A_2$ is in a pure state. 
Within the quasiparticle picture, the fact that $A_2=\bar A_1$ is in a pure state 
is irrelevant in~\eqref{n-mi}, because entanglement is generated locally. 
Therefore, we conclude that in the space-time scaling limit Eq.~\eqref{n-mi} holds true in general, i.e., also when 
$A_1\cup A_2$ is in a mixed state. 
(Away from the scaling limit, subleading, i.e. ${\mathcal O}(1)$ in $t$ and $\ell$, deviations  are expected.) 
This implies that also for two non-complementary intervals $\varepsilon^{(f/b)}(k)$ is equal to the analogous quantity for the 1/2-R\'enyi entropy. 

Eq.~\eqref{quasi-1} is the main result of this manuscript which is valid both for {\it interacting and free integrable models}.
Unfortunately, for interacting systems we still do not know how to 
extract  unambiguously the density of R\'enyi entropy in 
quasimomentum space, in spite of several recent advances~\cite{AlCa17,AlCa17b,mestyan-2018}.
For this reason, in the following we focus on free (bosonic and fermionic) models for which Eq.~\eqref{quasi-1} is still a very remarkable 
prediction (similar exact results in equilibrium are not available). 
We anyhow stress that the proportionality \eqref{n-mi} between negativity and R\'enyi mutual information (in the scaling limit) is a highly non trivial prediction 
that may be checked by tensor network techniques. 
We expect such a relation to break down for non-integrable models since it is crucially related to the existence of infinitely-living quasiparticle excitations, although 
the linear increase in time of the entanglement entropy of one interval remains valid even for ergodic
systems \cite{nahum-17,dmcf-06,lc-08,hk-13,fc-15,kctc-17,cdc-17,r-2017,ckt-18,ctd-18,bhy-17,evdcz-18}.

{\it Free models}.
For quenches in free models, 
the GGE describing the steady-state is identified 
by the quasimomenta occupations $\rho(k)$. 
A straightforward calculation gives the GGE 
R\'enyi entropies as \cite{peschel-2009,AlCa17}
\begin{equation}
\label{renyi-free}
\frac{S_{\textrm{GGE}}^{(\alpha)}}L=\pm\frac{1}{1-\alpha}\int_{-\pi}^\pi \frac{dk}{2\pi}\ln[\pm\rho(k)^\alpha+(1\mp\rho(k))^\alpha]. 
\end{equation}
Here the overall plus and minus signs are for fermionic and bosonic systems, respectively. 
The momentum integration is on $[-\pi,\pi]$ because we focus on lattice models.
Hence, our prediction is~\eqref{quasi-1} in which 
\begin{equation}
\varepsilon^{(f/b)}(k)=\pm\ln[\pm\rho(k)^{1/2}+(1\mp\rho(k))^{1/2}].
\end{equation} 
{\it A bosonic system}.
As an example of bosonic systems to numerically verify our results, we discuss 
quantum quenches in the harmonic chain, which describes a 
system of $L$ harmonic oscillators on a ring. Its hamiltonian 
reads $H=1/2\sum_{n=0}^{L-1} \left[ p_n^2+ m^2 q_n^2+ 
(q_{n+1}-q_n)^2\right]$,
where $q_n$ and $p_n$ are canonically conjugated variables, 
with $[q_n,p_m]=i\delta_{nm}$, and $m$ is a mass parameter. 
The harmonic chain is diagonalized by 
Fourier transform. The single particle energies are $e(k)=[m^2+2(1-\cos k)]^{1/2}$ and the group velocities $v(k)=de(k)/dk$. 
Here we consider a generic mass quench $m_0\to m$. 
The occupation density $\rho(k)$ describing the steady state is  \cite{cc-07}
\begin{equation}
\rho(k)=\frac{1}{4}\Big(\frac{e(k)}{e_0(k)}+\frac{e_0(k)}{e(k)}
\Big)-\frac{1}{2}, 
\end{equation}
with $e_0(k)$ and $e(k)$ being respectively the pre- and post-quench dispersions.

For bosonic systems the mutual information and the negativity~\eqref{neg} can be calculated 
from the two-point correlation function~\cite{audenaert-2002}. 
The results are shown in Fig.~\ref{fig1}. 
Panels (a) and (b) show the bosonic negativity for adjacent and disjoint intervals, respectively.  
Data are for several values of the intervals length $\ell$ up to $\ell\le 200$. Since we are interested 
in the scaling limit, we plot ${\cal E}_{A_1:A_2}^{\scriptscriptstyle(b)}/\ell$ 
versus the rescaled time $t/\ell$. For two adjacent intervals,  
the negativity exhibits a linear growth for $t/\ell\approx 1.25$, 
which reflects the maximum velocity being $v_{\textrm{max}}\approx 0.4$. 
For larger times we observe a slow decay toward zero for $t/\ell\to\infty$.
This slow decay  is due to the slower quasiparticles with $v<v_{\textrm{max}}$, as for the mutual information~\cite{alba-2018}. 
The dashed-dotted line is the theoretical prediction. 
Clearly, although finite-size corrections are present, the numerical 
data approach the theory prediction in the scaling limit. 

The case of two disjoint intervals is reported in Fig.~\ref{fig1} (b), showing data for 
$\ell\le 200$ and fixed ratio $d/\ell=1/5$. 
Now the negativity is zero up to $t=d/(2v_{\textrm{max}})$, as expected. Scaling corrections 
are smaller as compared with the case of adjacent intervals, and already the data 
for $\ell=200$ are indistinguishable from the scaling limit result (dashed-dotted line). 
This may be related to the property that stationary correlations decay exponentially with the distance.

{\it A fermionic system}.
To check our result for free fermions we consider a quantum quench in the transverse field Ising 
model (TFIM) with periodic boundary conditions, 
which is defined by the hamiltonian 
$H=-1/2\sum_{j=0}^{L-1}[\sigma_n^x\sigma_{n+1}^x+h\sigma_n^z]$. 
Here $\sigma_n^{x,z}$ are the Pauli matrices, and $h$ is the 
magnetic field. The Ising chain is mapped to free fermions by a Jordan-Wigner and 
a Bogoliubov transformation. The single particle energies read 
$e(k)=[h^2-2h\cos k+1]^{1/2}$. We consider the magnetic field 
quench $h_0\to h$. The quench is parametrized by 
the Bogoliubov angle $\cos\Delta(k)=(1+h h_0-(h+h_0)\cos k)/(e(k)e_0(k))$. 
The GGE density $\rho(k)$  is \cite{sps-04,cef-12}
\begin{equation}
\rho(k)=\frac{1}{2}(1-\cos\Delta(k)). 
\end{equation}

Unlike the bosonic case, the  negativity \eqref{neg} cannot be derived with correlation matrix techniques. 
The main problem is that the partial transpose 
in~\eqref{neg} is not a gaussian operator, but it can 
be written as the sum of two gaussian (non-commuting) operators $O_\pm$ as~\cite{eisler-2014a} 
\begin{equation}
\rho^{T_2}_{A}=\frac{1-i}{2} O_++\frac{1+i}{2}O_-. 
\end{equation}
Then, it is not immediate to calculate the spectrum of $\rho_{A}^{\scriptscriptstyle T_2}$ \cite{fc-10}
and the negativity thereof. 
For this reason the fermionic negativity has been introduced and it reads~\cite{shapourian-2016} 
\begin{equation}
\label{fneg}
{\cal E}^{(f)}_{A_1:A_2}=\ln\textrm{Tr}\sqrt{O_+O_-}. 
\end{equation}
It has been shown that for fermionic system ${\cal E}_{A_1:A_2}^{\scriptscriptstyle(f)}$ is an entanglement monotone. 
Moreover, as the product $O_+O_-$ in~\eqref{fneg} is 
a gaussian fermionic operator, ${\cal E}^{\scriptscriptstyle (f)}_{A_1:A_2}$ 
can be efficiently computed.

Numerical results for ${\cal E}_{A_1:A_2}^{\scriptscriptstyle(f)}$ 
for two adjacent intervals are discussed in Fig.~\ref{fig1} (c), 
for the quench with $h_0=10$ and $h=2$. 
The data exhibit a linear behavior up to $t/\ell\approx 0.5$, 
reflecting that $v_{\textrm{max}}\approx 1$.  
Similar to the bosonic case (see Fig.~\ref{fig1} (a)), finite-size corrections 
are present, although the theory result is recovered in the scaling limit. 
The  case of two disjoint intervals is presented in Fig.~\ref{fig1} (c), for geometries  
with fixed $d/\ell=1/20$ and $\ell\le 200$. Small deviations 
from the scaling-limit results are present, especially near the peak at 
$t/\ell\sim0.5$. 

{\it Finite-size corrections}.
\revision{In} Figure~\ref{fig1} (e) we investigate the corrections to the scaling, for both the fermionic and bosonic negativity. 
We focus on  adjacent intervals (Fig.~\ref{fig1} (a) and (c)), for which corrections are larger. The square and diamond symbols 
are the data for the bosonic and fermionic negativity at fixed $t/\ell=1.5$ and 
$t/\ell=0.5$, respectively. The $x$-axis shows $1/\ell$. 
The crosses are the theoretical results in the scaling limit. The dotted lines are fits to the behavior $1/\ell$, and are clearly consistent with the data. 
As a final check, for all considered quenches, we also computed the R\'enyi mutual information of order $1/2$ finding that in the space-time scaling limit
converges to the same curve as the negativity. We do not report the numerical data in Fig. \ref{fig1} in order to make it readable.

{\it Conclusions}. We provided a quantitative quasiparticle description for the dynamics of the logarithmic negativity after 
a quantum quench in integrable systems. 
Our main results are Eqs. \eqref{quasi-1} and \eqref{n-mi}. 
The latter establish an exact proportionality between the R\'enyi mutual information and the negativity that is expected to be valid for 
generic integrable models, even beyond the (standard) assumption of an initial state generating only pairs of quasiparticles 
(e.g. also in the cases considered in \cite{btc-18,bc-18}).
Eq. \eqref{quasi-1} is also generically valid (within the pairs assumption), but we can determine the single particle contribution to the negativity $\varepsilon^{(f/b)}(k)$
only for free models, because of well-known difficulties in obtaining the quasimomentum-space density of R\'enyi entropies \cite{AlCa17}.

Our work opens several research directions. 
First, for free fermionic models the techniques of Ref.~\cite{fagotti-2008} may be adapted 
to derive {\it ab initio} Eq.~\eqref{quasi-1}.
Second, \revision{our} result further motivates to derive the complete quasiparticle picture for the 
R\'enyi entropies. 
Concomitantly, it would be worth providing a thorough numerical analysis of 
the validity  of~\eqref{n-mi} in interacting models 
by using the tensor network methods, although \revision{it} is unclear whether the accessible times would probe the scaling regime. 
 An interesting direction would be to investigate the 
behavior of the negativity in long-range models~\cite{palencia1,palencia2}. 

Finally, there is no reason why \eqref{n-mi} should hold true in ergodic systems, 
as there is no notion of quasiparticles. The most likely scenario is that the negativity of disjoint intervals vanishes in the space-time scaling limit, reflecting 
the behavior suggested for the mutual information \cite{alba-2016,alba-2018, nahum-17,hol}. 
However, it would be enlightening to understand whether there is still a relation between these two very different quantities. 

\begin{acknowledgments}
{\it Acknowledgments.} 
PC acknowledges support from ERC under Consolidator grant  number 771536 (NEMO). 
VA acknowledges support from the European Union's Horizon 2020 under the Marie Sklodowska-Curie grant agreement No 702612 OEMBS.
Part of this work has been carried out during the workshop ``Quantum Paths''  at  the  Erwin  Schr\"odinger  International Institute
(ESI) in Vienna,  during the workshop ``Entanglement in quantum systems'' at the Galileo Galilei Institute (GGI) in Florence and 
during the workshop ``The Dynamics of Quantum Information'' at KITP. 

\end{acknowledgments}

\section{Additional data for the R\'enyi mutual information}

\begin{figure}
\centerline{\includegraphics[width=.9\linewidth]{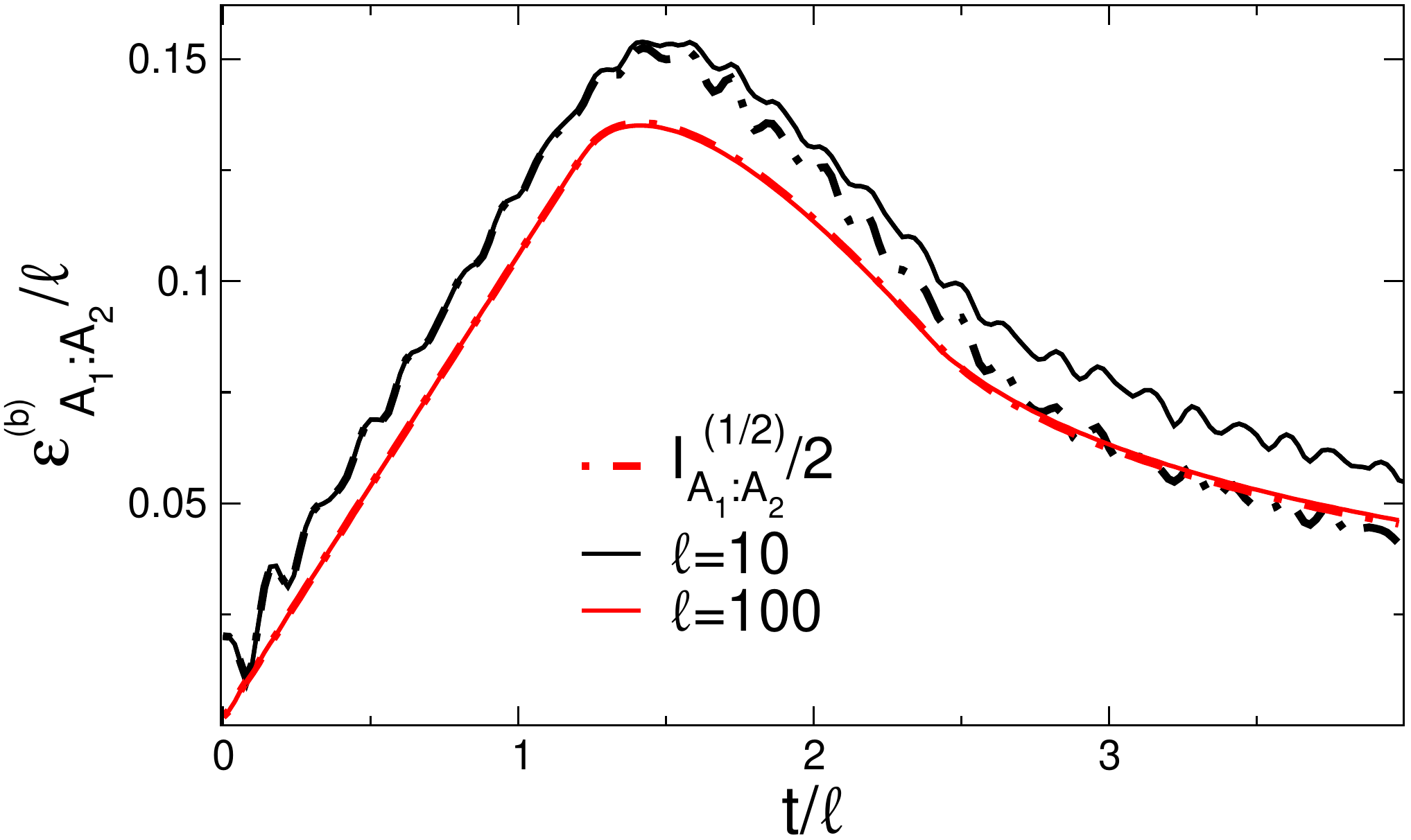}}
\caption{ Comparison between the bosonic negativity ${\cal E}_{A_1:A_2}^{(b)}$ 
 and the R\'enyi mutual information 
 $I_{A_1:A_2}^{(1/2)}$. The data are for two adjacent equal interals of 
 length $\ell$ after the mass quench in the harmonic chain (same as 
 in Fig.~\ref{fig1}). 
 The continuous and the dashed-dotted lines are the results for the negativity and 
 for the mutual information respectively. 
}
\label{fig_app}
\end{figure}

In this appendix we provide some numerical evidence 
supporting the validity of~\eqref{n-mi}. We consider 
the same mass quench as in Fig.~\ref{fig1} (a,b). 
In Fig.~\ref{fig_app} we focus on the bosonic mutal information 
${\cal E}_{A_1:A_2}^{(b)}$ between two interval of the 
same length $\ell$. The two intervals are adjacent. 
The continuous lines in the Figure are the 
results for the bosonic negativity for $\ell=10,100$. 
The dashed-dotted lines 
are the results for half of the R\'enyi mutual information with 
$\alpha=1/2$. As it is clear from the figure, at small values 
of $t/\ell$, Eq.~\eqref{n-mi} is verified already 
for $\ell=10$. At large values of $t/\ell$ deviations from 
Eq.~\eqref{n-mi} are visible. These, however, are due to 
the finite length of the interval. This is confirmed by the 
data for $\ell=100$. Now Eq.~\eqref{n-mi} holds up to 
$t\approx 3$. Finally, we should mention that similar result 
can be obtained for two disjoint intervals and for quenches in 
the Ising chain.

\end{document}